\documentclass[pra,twocolumn,superscriptaddress,showpacs,preprintnumbers,amsmath,amssymb]{revtex4}
\usepackage{mathrsfs}
\usepackage{bbm}
\usepackage{amsfonts}
\usepackage{tipa}

\usepackage{epsfig,graphicx}
\usepackage{amstext}
\usepackage{amsmath}
\usepackage{graphicx}
\usepackage{times}

\begin{document}
%%%%%%%%%%%%%%%%%%%%%%%%%%%%%%%%%%%%%%%%%%%%%%%%%%%%%%%%%%%%%%%%%%%%%%
%TCIDATA{OutputFilter=Latex.dll}
%TCIDATA{Version=5.00.0.2552}
%TCIDATA{<META NAME="SaveForMode" CONTENT="1">}
%TCIDATA{LastRevised=Wednesday, June 22, 2005 16:21:09}
%TCIDATA{<META NAME="GraphicsSave" CONTENT="32">}

\title{Evolution equation for geometric quantum correlation measures}

\author{Ming-Liang Hu}
\email{mingliang0301@163.com}
\affiliation{School of Science, Xi'an University of Posts and Telecommunications,
             Xi'an 710121, China}
\author{Heng Fan}
\email{hfan@iphy.ac.cn}
\affiliation{Beijing National Laboratory for Condensed Matter Physics,
             Institute of Physics, Chinese Academy of Sciences, Beijing 100190, China}
\affiliation{Collaborative Innovation Center of Quantum Matter, Beijing 100190, China}

\begin{abstract}
A simple relation is established for the evolution equation of
quantum information processing protocols such as quantum
teleportation, remote state preparation, Bell-inequality violation
and particularly dynamics of the geometric quantum correlation
measures. This relation shows that when the system traverses the
local quantum channel, various figures of merit of the quantum
correlations for different protocols demonstrate a factorization
decay behavior for dynamics. We identified the family of quantum
states for different kinds of quantum channels under the action of
which the relation holds. This relation simplifies the assessment of
many quantum tasks.

\end{abstract}


\maketitle

\section{Introduction}\label{sec:1}
Quantum theory enables the processing of quantum information with
efficiency that outperforms the classical information processing.
The quantum correlations, quantified by various measures, of the
computational systems were considered to be the origin of this
superiority \cite{RMP1}. But when implementing the quantum
information processing (QIP) tasks experimentally, one has to face
the unavoidable interaction of the quantum devices with their
environments, the process of which often induces decay of
correlations because of decoherence \cite{open}. The lost of quantum
correlation can generally damage the superiority of QIP. Therefore,
it is of practical significance to make clear the behavior of a
correlation when it serves as a resource in carrying out certain QIP
tasks. The process of the decoherence effects can be in general
described as the time evolution of the quantum system.

To assess the time evolution of a given quantum correlation measure,
one usually needs to derive first the time-dependence of the density
matrix, and this is intractable. From a theoretical point of view,
one needs to obtain the evolution of the system for every initial
states, while from an experimental point of view, it requires the
full reconstruction of the evolved state via quantum tomography
\cite{tomography}. This either needs much computational resources or
is hard to do experimentally.

Seeking general dynamical law for various correlation measures can
simplify these tasks. As an earlier proposed scenario of quantum
correlation measure, entanglement has been the focus of researchers
for years \cite{RMP1}. Particularly, when one party of a two-qubit
system traverses a noisy channel, the evolution equation of
concurrence \cite{con} was found to be governed by a factorization
law for the initial pure states \cite{natphys}. This simplifies the
assessment of the evolution of concurrence as to evaluate a decay
factor. Soon a universal curve describing the evolution of
concurrence for general two-qubit states was revealed
\cite{science}. Since then, the evolution equations for many other
entanglement measures \cite{FL1,FL2,FL3} or their bounds
\cite{FLb1,FLb2,FLb3} were derived.

Besides entanglement, the quantumness of correlations can also be
characterized from other perspectives. Those measures are such as
the conventional quantum discord \cite{qd}, geometric quantum discord
(GQD) \cite{gqd,gqd1,trace,trace1,lqu,square}, and measurement-induced
nonlocality \cite{min,min1}, which are also resources for certain
QIP tasks \cite{appl,Dakic,Guml}. Moreover, many measures related to
the explicit quantum tasks (see the following text) reveal also some
kinds of correlations. The decay dynamics for them under different
environments have been extensively studied \cite{Modirmp,dyn0,dyn1,
dyn2,dyn3,dyn4,dyn5}. Yet, some general conclusions still remain
absent. Motivated by these, we study in this paper the evolution
equations for various geometric correlation measures when the system
traverses a local quantum channel. We will show that for a broad
class of quantum states, the evolution equations of the considered
correlations are fully characterized by the product of the initial
correlation and a time-dependent factor determined solely by the
structure of the channel. This facilitates greatly the estimation of
the robustness of the correlations in realistic physical systems.

\section{Preliminaries}\label{sec:2}
Consider a general bipartite state $\rho$ in the Hilbert space
$\mathcal {H}_{ab}$. It can always be decomposed as
%%%%%%%%%%%%%%%%%%%%%%%%%%%
\begin{eqnarray}\label{eq1}
 \rho = -\frac{1}{d_a d_b} \mathbb{I}_a\otimes \mathbb{I}_b
        +\rho_a\otimes \frac{1}{d_b}\mathbb{I}_b+\frac{1}{d_a}
        \mathbb{I}_a\otimes\rho_b+\rho_c,
\end{eqnarray}
%%%%%%%%%%%%%%%%%%%%%%%%%%%
where the reduced states $\rho_a = {\rm tr}_b \rho$, $\rho_b={\rm
tr}_a \rho$, and the traceless `correlation operator' $\rho_c$ are
give by
%%%%%%%%%%%%%%%%%%%%%%%%%%%
\begin{equation}\label{eq2}
 \begin{split}
  & \rho_a = \frac{1}{d_a}\mathbb{I}_a+\vec{x}\cdot\vec{X},~
    \rho_b = \frac{1}{d_b}\mathbb{I}_b+\vec{y}\cdot\vec{Y},\\
  & \rho_c = \sum_{i=1}^{d_a^2-1}\sum_{j=1}^{d_b^2-1}t_{ij} X_i \otimes Y_j,
 \end{split}
\end{equation}
%%%%%%%%%%%%%%%%%%%%%%%%%%%
with $\vec{x}=(x_1,x_2,\ldots,x_{d_a^2-1})^{t}$, $\vec{X}=(X_1,X_2,
\ldots,X_{d_a^2-1})^{t}$ ($d_{a}={\rm dim}\mathcal {H}_{a}$, and $t$
denotes transpose), and likewise for $\vec{y}$ and $\vec{Y}$.
$\{\mathbb{I}_a/\sqrt{d_a}, \vec{X}\}$ and $\{\mathbb{I}_b /
\sqrt{d_b},\vec{Y}\}$ constitute the orthonormal operator bases for
$\mathcal {H}_a$ and $\mathcal {H}_b$, respectively.

Two extensively studied cases in the literature are $d_{a,b}=2$ and
$3$, for which $X_i$ (and $Y_j$) are given respectively by $\sigma_i
/\sqrt{2}~$ ($i=1,2,3$) and $\lambda_j/ \sqrt{2}~$ ($j=1,2,\ldots,
8$), with $\sigma_i$ and $\lambda_j$ being the Pauli and the
Gell-Mann matrices. They can describe the qubit, the qutrit, and the
hybrid qubit-qutrit systems, which are of central relevance to QIP.

The decomposed $\rho$ in Eq. \eqref{eq1} enables the definition of
correlation measures from a geometric perspective. We consider here
the definition of the general form
%%%%%%%%%%%%%%%%%%%%%%%%%%%
\begin{eqnarray}\label{eq3}
 D_p(\rho) = \mathop{\rm opt}_{\Pi^a\in \mathcal {M}}
             \parallel\rho-\Pi^a(\rho)\parallel_p^p,
\end{eqnarray}
%%%%%%%%%%%%%%%%%%%%%%%%%%%
where $\parallel\chi\parallel_p=[\mbox{tr}(\chi^\dag\chi)^{p/2}]
^{1/p}$ is the Schatten $p$-norm, and opt represents the
optimization over some class $\mathcal {M}$ of the local
measurements $\Pi^a=\{\Pi_k^a\}$ acting on party $a$.

Eq. \eqref{eq3} covers a series of discord-like correlation measures
being proposed recently. (i) If opt represents minimum and $\mathcal
{M}$ is that of the projective measurements, one recovers the 2-norm
GQD for $p=2$ \cite{gqd}, and the 1-norm GQD for $p=1$ \cite{trace}.
(ii) If $\rho$ is replaced by $\sqrt{\rho}$, then one obtains the
Hellinger distance quantum discord for $p=2$ \cite{lqu,square}.
(iii) If opt represents maximum and $\mathcal {M}$ is confined to
the locally invariant measurements that maintain $\rho_a$, Eq.
\eqref{eq3} turns to be the conventional measurement-induced
nonlocality for $p=2$ \cite{min}, and its modified version for $p=1$
\cite{min1}.

Besides $D_p (\rho)$ that is determined by $\vec{x}$, $\vec{y}$, and
the matrix $T = (t_{ij})$, there are other measures related to
explicit quantum protocols that are determined solely by $T$. We
consider here the protocols of quantum teleportation, remote state
preparation, and Bell-inequality violation. It has been shown that
the average fidelity of quantum teleportation \cite{fide}, remote
state preparation \cite{Dakic}, and the maximum Bell-inequality
violation \cite{Bell} were given respectively by
%%%%%%%%%%%%%%%%%%%%%%%%%%
\begin{equation}\label{eq4}
 \begin{split}
  & \mathcal {F}_{\rm qt}(\rho) = \frac{1}{2}+\frac{1}{6}
                                  \mathcal{N}_{\rm qt}(\rho),~
    \mathcal {F}_{\rm rsp}(\rho)= \frac{1}{2}(E_2 + E_3), \\
  & \mathcal {B}_{\rm max}(\rho)= 2\sqrt{E_1 + E_2},
 \end{split}
\end{equation}
%%%%%%%%%%%%%%%%%%%%%%%%%%%
where $ \mathcal {N}_{\rm qt}(\rho) = {\rm tr}\sqrt{T^\dag T}$, and
$E_1 \geq E_2 \geq E_3$ are the eigenvalues of $T^\dag T$.

\section{General results}\label{sec:3}
We restrict ourselves in the following to the projective
measurements $\Pi^a$, then it follows from Eqs. \eqref{eq1} and
\eqref{eq3} that $D_p (\rho)$ is solely determined by $\vec{x}$ and
$T$, i.e.,
%%%%%%%%%%%%%%%%%%%%%%%%%%%
\begin{equation}\label{eq-new1}
 D_p(\rho)= \mathop{\rm opt}_{\Pi^a\in \mathcal {M}}
            \parallel \varrho-\Pi^a(\varrho)\parallel_p^p,
\end{equation}
%%%%%%%%%%%%%%%%%%%%%%%%%%%
where
%%%%%%%%%%%%%%%%%%%%%%%%%%%
\begin{equation}\label{eq-new2}
 \varrho = \vec{x} \cdot \vec{X} \otimes
           \frac{1}{d_b}\mathbb{I}_b+\rho_c.
\end{equation}
%%%%%%%%%%%%%%%%%%%%%%%%%%%

\begin{figure}
\centering
\resizebox{0.31\textwidth}{!}{%
\includegraphics{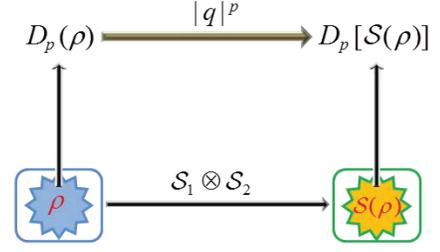}}
% If not, use\vspace{5cm}
% Give the correct figure height in cm
\caption{(Color online) Comparison of state evolution and geometric
         correlation dynamics. For certain family of $\rho$ (denoted
         by the color shaded region) the evolution equation for
         $D_p[\mathcal {S}(\rho)]$ can be deduced from $D_p (\rho)$.} \label{fig:1}
% Give a unique label
\end{figure}
%%%%%%%%%%%%%%%%%%%%%%%%%%%%%%

Now, we suppose the considered system $ab$ passes through a quantum
channel $\mathcal {S} $ such that
%%%%%%%%%%%%%%%%%%%%%%%%%%%
\begin{eqnarray}\label{eq5}
 \vec{x}' = q(t)\vec{x},~  T' = q(t) T,
\end{eqnarray}
%%%%%%%%%%%%%%%%%%%%%%%%%%%
with $q(t)$ being a time-dependent factor that contains only the
information on the channel's structure, and the primed parameters
$\vec{x}'$ and $T'$ denote respectively the local Bloch vector and
the correlation tensor for $\mathcal {S}(\rho)$, similar as those
for $\rho$ in Eq. \eqref{eq2}. Then, from Eq. \eqref{eq5} we know
that $\varrho'= q(t)\varrho$, and therefore, as illustrated in Fig.
\ref{fig:1}, the formula of Eq. \eqref{eq-new1} implies that the
evolution equation of $D_p[\mathcal {S}(\rho)]$ fulfills the
following factorization decay behavior
%%%%%%%%%%%%%%%%%%%%%%%%%%%
\begin{eqnarray}\label{eq6}
 D_p[\mathcal {S}(\rho)] = |q(t)|^p D_p(\rho),
\end{eqnarray}
%%%%%%%%%%%%%%%%%%%%%%%%%%%
which is solely determined by the product of the initial $D_p(\rho)$
and a channel-dependent factor $|q(t)|$.

In the operator-sum representation, the evolved state of the system
$ab$ under the action of the local quantum channel $\mathcal {S}_1
\otimes \mathcal {S}_2$ (it reduces to the one-sided channel when
$\mathcal {S}_1=\mathbb{I}_a$ or $\mathcal {S}_2 = \mathbb{I}_b$)
can be written compactly as
%%%%%%%%%%%%%%%%%%%%%%%%%%%
\begin{equation}\label{eq-new3}
 (\mathcal {S}_1 \otimes \mathcal {S}_2)\rho = \sum_{\mu\nu} E_{\mu\nu}\rho
                           E_{\mu\nu}^\dag,
\end{equation}
%%%%%%%%%%%%%%%%%%%%%%%%%%%
where $\mathcal {S}_1$ denotes the channel for subsystem $a$ and
$\mathcal {S}_2$ for subsystem $b$, and the Kraus operators
$E_{\mu\nu}=E_\mu \otimes E_\nu$, with $E_\mu$ and $E_\nu$ satisfy
$\sum_{\mu}E_{\mu}^\dag E_{\mu} = \mathbb{I}_{a}$ and $\sum_{\nu}
E_{\nu}^\dag E_{\nu} = \mathbb{I}_{b}$.

In order to obtain the conditions under which the evolution
equations of $D_p[\mathcal {S}(\rho)]$ obeying the relation
\eqref{eq6}, we turn to the Heisenberg picture. Then as $x_i = {\rm
tr}\rho (X_i \otimes \mathbb{I}_b)$, $y_j = {\rm tr}\rho
(\mathbb{I}_a \otimes Y_j)$, and $t_{ij} = {\rm tr} \rho(X_i \otimes
Y_j)$ from Eq. \eqref{eq1}, we obtain
%%%%%%%%%%%%%%%%%%%%%%%%%%%
\begin{eqnarray}\label{eq-sp11}
 x_i' = {\rm tr}[\rho \mathcal {S}_1^\dag(X_i) \otimes \mathbb{I}_b ],~
 y_j' = {\rm tr}[\rho \mathbb{I}_a \otimes \mathcal {S}_2^\dag (Y_j)],
\end{eqnarray}
%%%%%%%%%%%%%%%%%%%%%%%%%%%
and
%%%%%%%%%%%%%%%%%%%%%%%%%%%
\begin{equation}\label{eq-sp12}
 \begin{split}
  t_{ij}' & = {\rm tr}[(\mathcal {S}_1\otimes \mathcal {S}_2)\rho(X_i\otimes Y_j)] \\
          & = \sum_{\mu\nu}{\rm tr}[E_{\mu\nu}\rho
              E_{\mu\nu}^\dag(X_i\otimes Y_j)] \\
          & = \sum_{\mu\nu}{\rm tr}[ \rho E_{\mu\nu}^\dag
              (X_i \otimes Y_j) E_{\mu\nu}] \\
          & = {\rm tr}[\rho \mathcal {S}_1^\dag(X_i)\otimes\mathcal {S}_2^\dag (Y_j)],
 \end{split}
\end{equation}
%%%%%%%%%%%%%%%%%%%%%%%%%%%
where $\mathcal {S}_1^\dag(X_i) = \sum_\mu E_\mu^\dag X_i E_\mu$
denotes the map of $\mathcal {S}_1$ on $X_i$, and likewise for
$\mathcal {S}_2^\dag(Y_j)$. We have also used the facts $\mathcal
{S}_1^\dag(\mathbb{I}_a) = \mathbb{I}_a$ and $\mathcal
{S}_2^\dag(\mathbb{I}_b) = \mathbb{I}_b$ when deriving Eq.
\eqref{eq-sp11}.

Then, from the above two equations one can derive that Eq.
\eqref{eq5} can be satisfied by a broad class of quantum channels,
for which we will discuss in the framework of the one-sided cases
$\mathcal {S}_1\otimes \mathbb{I}_b$ and $\mathbb{I}_a\otimes
\mathcal {S}_2 $, and the two-sided case $\mathcal {S}_1
\otimes\mathcal {S}_2$.

\emph{Theorem 1.} If a channel gives $\mathcal {S}_1^\dag (X_i)=q_a
X_i$ for all $\{X_i\}$, and $\mathcal {S}_2^\dag (Y_j) = q_b Y_j$
for all $\{Y_j\}$, then $D_p[\mathcal {S}(\rho)]$ obeys the
factorization decay behavior of Eq. \eqref{eq6} for the families of
$\rho$ with
%%%%%%%%%%%%%%%%%%%%%%%%%%%
\begin{equation}\label{eq-c1}
 \begin{aligned}
  (1)&~ \mbox{arbitrary}~ \rho_a,~ \rho_b,~ \rho_c ~~
        (\mbox{for \,} \mathcal {S}_1 \otimes \mathbb{I}_b), \\
  (2)&~ \rho_a = \frac{1}{d_a}\mathbb{I}_a, \mbox{or~} \rho_c=0 ~~
        (\mbox{for \,} \mathcal {S}_1 \otimes \mathcal {S}_2~ \mbox{with~} \mathcal {S}_2\neq \mathbb{I}_b).
 \end{aligned}
\end{equation}
%%%%%%%%%%%%%%%%%%%%%%%%%%%

\emph{Proof.} As for $\mathcal {S}_1 \otimes \mathcal {S}_2$, we
have $\vec{x}'= q_a \vec{x} $ and $T'=q_a q_b T$. Then it is direct
to see that for the special case $\mathcal {S}_2=\mathbb{I}_b$ which
corresponds to $q_b=1$, Eq. \eqref{eq5} is always satisfied, and
thus the factorization decay behavior of Eq. \eqref{eq6} holds for
arbitrary bipartite state $\rho$, i.e., there is no restriction on
$\rho_a$, $\rho_b$, and $\rho_c$. For $\mathcal {S}_1 \otimes
\mathcal {S}_2$ with $\mathcal {S}_2\neq \mathbb{I}_b$, however, Eq.
\eqref{eq5} is satisfied only when $\vec{x}=0$ or $T=0$, which
corresponds to the family of $\rho$ with $\rho_a =
\mathbb{I}_a/{d_a}$ or $\rho_c =0$. This completes the proof.
\hfill{$\blacksquare $}

In general, $\mathcal {S}_1 \otimes \mathcal {S}_2$ includes the
channels $\mathcal {S}_1 \otimes \mathbb{I}_b$ and $\mathbb{I}_a
\otimes \mathcal {S}_2$ as two special cases, and thus the range of
states for $\mathcal {S}_1 \otimes \mathbb{I}_b$ and $\mathbb{I}_a
\otimes \mathcal {S}_2$ should be at least as large as that for
$\mathcal {S}_1 \otimes \mathcal {S}_2$. This is indeed what Theorem
1 implies, as there is no restriction on the range of states for
family (1), while family (2) includes only those $\rho$ with $\rho_a
= \mathbb{I}_a / {d_a}$ or $\rho_c =0$.

\emph{Theorem 2.} If $\mathcal {S}_1^\dag (X_k) = q_a X_k$ only for
$\{X_k\}$ with $k = \{k_1,\ldots,k_\alpha\}$ ($\alpha < d_a^2-1$),
and $\mathcal {S}_2^\dag (Y_l) = q_b Y_l$ only for $\{Y_l\}$ with
$l=\{l_1,\ldots,l_\beta\}$ ($\beta < d_b^2-1$), then $D_p[\mathcal
{S}(\rho)]$ obeys the factorization decay behavior of Eq.
\eqref{eq6} for the families of $\rho$ with
%%%%%%%%%%%%%%%%%%%%%%%%%%%
\begin{equation}\label{eq-c2}
 \begin{aligned}
  (1)&~ \rho_a = \rho_a^{(1)},~ \rho_c = \rho_c^{(1)} ~~
        (\mbox{for \,} \mathcal {S}_1 \otimes \mathbb{I}_b),\\
  (2)&~ \rho_a = \frac{1}{d_a}\mathbb{I}_a,~ \rho_c = \rho_c^{(2)},~
        \mbox{or~} \rho_c = 0~
        (\mbox{for \,} \mathbb{I}_a \otimes \mathcal {S}_2),\\
  (3)&~ \rho_a = \frac{1}{d_a}\mathbb{I}_a,~ \rho_c = \rho_c^{(3)},~
        \mbox{or~} \rho_a = \rho_a^{(1)},~\rho_c = 0 \\
     &~ (\mbox{for \,} \mathcal {S}_1 \otimes \mathcal {S}_2),
 \end{aligned}
\end{equation}
%%%%%%%%%%%%%%%%%%%%%%%%%%%
where
%%%%%%%%%%%%%%%%%%%%%%%%%%%
\begin{equation}\label{eq-cc12}
 \begin{split}
  & \rho_a^{(1)} = \frac{1}{d_a}\mathbb{I}_a +
                   \sum_{k=k_1}^{k_\alpha} x_k X_k,~
    \rho_c^{(1)} = \sum_{k=k_1}^{k_\alpha}\sum_{j=1}^{d_b^2-1}
                   t_{k j} X_k \otimes Y_j,\\
  & \rho_c^{(2)} = \sum_{i=1}^{d_a^2-1}\sum_{l=l_1}^{l_\beta}
                   t_{i l} X_i \otimes Y_l,~
    \rho_c^{(3)} = \sum_{k=k_1}^{k_\alpha}\sum_{l=l_1}^{l_\beta}
                   t_{k l} X_k \otimes Y_l.
 \end{split}
\end{equation}
%%%%%%%%%%%%%%%%%%%%%%%%%%%

\emph{Proof.} As $\mathcal {S}_1^\dag (X_k)=q_a X_k$ and $\mathcal
{S}_2^\dag (Y_l)=q_b Y_l$ only for partial $\{X_k\}$ and $\{Y_l\}$,
then if components of the local Bloch vector and the correlation
tensor, namely, those $x_i$ and $t_{ij}$ related to $\mathcal
{S}_1^\dag (X_i) \neq q_a X_i$ or/and $\mathcal {S}_2^\dag (Y_j)
\neq q_b Y_j$ equal zero, the requirement in Eq. \eqref{eq5} is
always satisfied for $\mathcal {S}_1 \otimes \mathbb{I}_b$, while it
is satisfied for $\mathbb{I}_a \otimes \mathcal {S}_2$ and $\mathcal
{S}_1 \otimes \mathcal {S}_2$ by further choosing $\vec{x}=0$ or
$T=0$. Thus, the families of $\rho$ for which the factorization
decay behavior of Eq. \eqref{eq6} holds can be written by combining
these with Eq. \eqref{eq2}, which are of the form of Eq.
\eqref{eq-c2}. \hfill{$\blacksquare $}

Moreover, from Eq. \eqref{eq4} one can show that the evolution
equations of the measures related to quantum teleportation, remote
state preparation, and Bell-inequality violation can also
demonstrate some factorization decay behaviors when the system $ab$
traverses a quantum channel such that $T'$ for $\mathcal {S}(\rho)$
is given by $T'=q(t) T$. They are
%%%%%%%%%%%%%%%%%%%%%%%%%%%
\begin{equation}\label{eq12}
\begin{split}
 & \mathcal {N}_{\rm qt}[\mathcal {S}(\rho)] = |q(t)|\mathcal {N}_{\rm qt}(\rho), \\
 & \mathcal {F}_{\rm rsp}[\mathcal {S}(\rho)] = |q(t)|^2\mathcal{F}_{\rm rsp}(\rho), \\
 & \mathcal {B}_{\rm max}[\mathcal {S}(\rho)] = |q(t)|\mathcal {B}_{\rm max}(\rho),
 \end{split}
\end{equation}
%%%%%%%%%%%%%%%%%%%%%%%%%%%
and after a similar analysis to that for proving Theorem 2, one can
show that the requirement $T'=q(t) T$ can be satisfied by the
families of $\rho$ with $\rho_c=\rho_c^{(1)}$ for $\mathcal
{S}_1\otimes \mathbb{I}_b$, $\rho_c=\rho_c^{(2)}$ for
$\mathbb{I}_a\otimes\mathcal {S}_2$, and $\rho_c=\rho_c^{(3)}$ for
$\mathcal {S}_1\otimes\mathcal {S}_2$.

Of course, the above conditions are sufficient but not necessary,
namely, for certain specifically defined correlation measures, the
relations in Eqs. \eqref{eq6} and \eqref{eq12} may hold even if
those requirements cannot be satisfied \cite{zhang}. The
significance of the above conditions lies in that they provide a
flexible way for identifying the families of $\rho$ for which the
evolution equations of the considered correlations demonstrate a
factorization decay behavior, and this is of practical significance
for assessing the robustness of certain quantum protocols.

In practice, one can determine if a given channel allows the
factorization decay behavior of quantum correlations by full
tomography of the channel, the procedure of which requires
measurements with a limited set of specific states, and thus is
experimentally feasible in principle \cite{tomography2}.

The factorization decay behavior presented above can also be
generalized to the symmetric version of the discord-like geometric
correlation measures \cite{Modirmp}
%%%%%%%%%%%%%%%%%%%%%%%%%%%
\begin{eqnarray}\label{eq34}
 \tilde{D}_p(\rho) = \mathop{\rm opt}_{\Pi^{ab} \in \mathcal {M}}
                     ||\rho-\Pi^{ab}(\rho)||_p^p,
\end{eqnarray}
%%%%%%%%%%%%%%%%%%%%%%%%%%%
where the optimization is now taken over the two-sided local
projection-valued measurements $\Pi^{ab} = \{\Pi_k^a \otimes
\Pi_l^b\}$. For this case, if the considered channel yields
%%%%%%%%%%%%%%%%%%%%%%%%%%%
\begin{eqnarray}\label{eq35}
 \vec{x}' = q(t)\vec{x},~\vec{y}' = q(t)\vec{y},~T'=q(t) T,
\end{eqnarray}
%%%%%%%%%%%%%%%%%%%%%%%%%%%
then by combining Eqs. \eqref{eq1}, \eqref{eq2}, and \eqref{eq34},
one can show that the evolution equation of $\tilde{D}_p[\mathcal
{S}(\rho)]$ still abides by a factorization decay behavior of the
form of Eq. \eqref{eq6}. The families of $\rho$ are as follows:

\emph{Theorem 3.} If $\mathcal {S}_1^\dag (X_i) = q_a X_i$ and
$\mathcal {S}_2^\dag (Y_j) = q_b Y_j$ for all $\{X_i\}$ and
$\{Y_j\}$, then the factorization decay behavior of
$\tilde{D}_p[\mathcal {S}(\rho)]$ holds for the families of $\rho$
with
%%%%%%%%%%%%%%%%%%%%%%%%%%%
\begin{equation}\label{eq-c3}
 \begin{aligned}
  (1)&~ \rho_b = \frac{1}{d_b}\mathbb{I}_b,~
        \mbox{or}~ \rho_a = \frac{1}{d_a}\mathbb{I}_a,~ \rho_c = 0 ~
        (\mbox{for \,} \mathcal {S}_1 \otimes \mathbb{I}_b),\\
  (2)&~ \rho_a = \frac{1}{d_a}\mathbb{I}_a, ~
        \mbox{or}~ \rho_b = \frac{1}{d_b}\mathbb{I}_b,~ \rho_c = 0 ~
        (\mbox{for \,} \mathbb{I}_a \otimes \mathcal {S}_2),\\
  (3)&~ \rho_a = \frac{1}{d_a}\mathbb{I}_a,~
        \rho_b = \frac{1}{d_b}\mathbb{I}_b,
        \mbox{or}~ \rho_a = \frac{1}{d_a}\mathbb{I}_a,~ \rho_c = 0,\\
     &~ \mbox{or}~ \rho_b = \frac{1}{d_b}\mathbb{I}_b,~ \rho_c = 0~~
        (\mbox{for \,} \mathcal {S}_1 \otimes \mathcal {S}_2).
 \end{aligned}
\end{equation}
%%%%%%%%%%%%%%%%%%%%%%%%%%%

\emph{Proof.} As we have $\vec{x}' = q_a \vec{x} $,
$\vec{y}'=q_b\vec{y}$, and $T'=q_a q_b T$ for general $\mathcal
{S}_1 \otimes \mathcal {S}_2$, Eq. \eqref{eq35} is satisfied when
$\vec{y}=0$, or $\vec{x}=0$ and $T=0$ for $\mathcal {S}_1 \otimes
\mathbb{I}_b$, and $\vec{x}=0$, or $\vec{y}=0$ and $T=0$ for
$\mathbb{I}_a \otimes \mathcal {S}_2$, which correspond to those
$\rho$ listed in families (1) and (2) of Eq. \eqref{eq-c3}. For
$\mathcal {S}_1 \otimes \mathcal {S}_2$, however, Eq. \eqref{eq35}
holds when $\vec{x} =\vec{y} =0$, or $\vec{x} =0$ and $T =0$, or
$\vec{y}=0$ and $T=0$. These correspond to the states listed in
family (3) of Eq. \eqref{eq-c3}. \hfill{$\blacksquare $}

Moreover, if $\mathcal {S}_1$ and $\mathcal {S}_2$ are the same, we
have $q_a=q_b$, and then for $\mathcal {S}_1\otimes\mathcal {S}_2$
the factorization decay behavior of $\tilde{D}_p [\mathcal
{S}(\rho)]$ also holds for the family of $\rho$ with only $\rho_c=
0$.

\emph{Theorem 4.} If $\mathcal {S}_1^\dag (X_k) = q_a X_k$ only for
$\{X_k\}$ with $k = \{k_1, \ldots,k_\alpha\}$ ($\alpha < d_a^2-1$),
and $\mathcal {S}_2^\dag(Y_l) = q_b Y_l$ only for $\{Y_l\}$ with $l=
\{l_1,\ldots, l_\beta\}$ ($\beta < d_b^2-1$), then the factorization
decay behavior of $\tilde{D}_p[\mathcal {S}(\rho)]$ holds for the
families of $\rho$ with
%%%%%%%%%%%%%%%%%%%%%%%%%%%
\begin{equation}\label{eq-c4}
 \begin{aligned}
  (1)&~ \rho_a = \rho_a^{(1)},~ \rho_b = \frac{1}{d_b}\mathbb{I}_b,~
                 \rho_c = \rho_c^{(1)},~ \mbox{or}~ \rho_a = \frac{1}{d_a}\mathbb{I}_a,\\
     &~ \rho_c = 0~ (\mbox{for \,} \mathcal {S}_1 \otimes \mathbb{I}_b),\\
  (2)&~ \rho_a = \frac{1}{d_a}\mathbb{I}_a,~ \rho_b = \rho_b^{(1)},~
                 \rho_c = \rho_c^{(2)},~ \mbox{or}~ \rho_b = \frac{1}{d_b}\mathbb{I}_b,\\
     &~ \rho_c = 0~ (\mbox{for \,} \mathbb{I}_a \otimes \mathcal {S}_2),\\
  (3)&~ \rho_a = \frac{1}{d_a}\mathbb{I}_a,~ \rho_b=\frac{1}{d_b}\mathbb{I}_b,~
                 \rho_c=\rho_c^{(3)},\\
     &~ \mbox{or~} \rho_\xi = \rho_\xi^{(1)},~ \rho_\zeta=\frac{1}{d_\zeta}
                   \mathbb{I}_\zeta,~ \rho_c=0~~
                   (\mbox{for \,} \mathcal {S}_1 \otimes \mathcal {S}_2),
 \end{aligned}
\end{equation}
%%%%%%%%%%%%%%%%%%%%%%%%%%%
where $\xi=a$ and $\zeta=b$, or $\xi=b$ and $\zeta=a$, with
%%%%%%%%%%%%%%%%%%%%%%%%%%%
\begin{equation}\label{eq-cc34}
 \rho_b^{(1)} =\frac{1}{d_b} \mathbb{I}_b +
               \sum_{l=l_1}^{l_\beta} y_l Y_l.
\end{equation}
%%%%%%%%%%%%%%%%%%%%%%%%%%%

\emph{Proof.} From the given conditions we know that Eq.
\eqref{eq35} is satisfied by choosing some of the parameters $x_i$,
$y_j$, and $t_{ij}$ to be zero. To be explicit, by choosing (1)
$\vec{y}$, $x_i$ and $t_{ij}$ related to $\mathcal {S}_1^\dag (X_i)
\neq q_a X_i$, or $\vec{x}$ and $T$ to be zero for $\mathcal {S}_1
\otimes \mathbb{I}_b$, (2) $\vec{x}$, $y_j$ and $t_{ij}$ related to
$\mathcal {S}_2^\dag (Y_j) \neq q_b Y_j$, or $\vec{y}$ and $T$ to be
zero for $\mathbb{I}_a \otimes \mathcal {S}_2$, and (3) $\vec{x}$,
$\vec{y}$ and $t_{ij}$, or $x_i$, $\vec{y}$, and $T$, or $\vec{x}$,
$y_j$, and $T$ related to $\mathcal {S}_1^\dag (X_i) \neq q_a X_i$
or/and $\mathcal {S}_2^\dag (Y_j) \neq q_b Y_j$ to be zero for
$\mathcal {S}_1 \otimes \mathcal {S}_2$. Then, by Eq. \eqref{eq2} it
is obvious that the families of $\rho$ are of the form of Eq.
\eqref{eq-c4}. \hfill{$\blacksquare $}

Moreover, if $\mathcal {S}_1$ and $\mathcal {S}_2$ are the same,
$q_a=q_b$, then under the action of $\mathcal {S}_1\otimes\mathcal
{S}_2$ the correlation measure $\tilde{D}_p [\mathcal {S}(\rho)]$
obeys the factorization decay behavior also for those $\rho$ with
$\rho_a = \rho_a^{(1)}$, $\rho_b = \rho_b^{(1)}$, and $\rho_c= 0$.

\section{Explicit Examples}\label{sec:4}
We construct in the following some quantum channels that satisfy the
conditions listed in the above theorems, and therefore the
factorization decay behavior presented in Eqs. \eqref{eq6} and
\eqref{eq12} are obeyed by a broad class of $\rho$.

\subsection{Depolarizing channel}
Consider first the depolarizing channel, which represents the
process in which a state $\rho_s$ is dynamically replaced by the
maximally mixed one. It gives
%%%%%%%%%%%%%%%%%%%%%%%%%%%
\begin{eqnarray}\label{eq13}
 \mathcal {S}_i(\rho_s) = q_s\rho_s+(1-q_s)\frac{1}{d_s}\mathbb{I}_s,
\end{eqnarray}
%%%%%%%%%%%%%%%%%%%%%%%%%%%
where $i = \{1,2\}$ and $s=\{a,b\}$. Eq. \eqref{eq13} corresponds to
the map $\mathcal {S}_1^\dag (X_i) = q_a X_i$ and $\mathcal
{S}_2^\dag(Y_j)=q_b Y_j$ for all $\{X_i\}$ and $\{Y_j\}$. Therefore,
the factorization decay behavior of Eq. \eqref{eq6} holds for
arbitrary $\rho$ if $\mathcal {S}_1 \otimes \mathbb{I}_b$ is
applied, while it holds for those $\rho$ with $\rho_a =
\mathbb{I}_a/d_a$ or $\rho_c = 0$ if $\mathbb{I}_a \otimes \mathcal
{S}_2$ and $\mathcal {S}_1 \otimes \mathcal {S}_2$ are applied.
Moreover, the relations in Eq. \eqref{eq12} hold for any two-qubit
state $\rho$, whether the depolarizing channel is one-sided or
two-sided.

The fact that the evolution equation of any $D_p[\mathcal
{S}(\rho)]$ obeys a factorization decay behavior of Eq. \eqref{eq6}
under the action of the depolarizing channel $\mathcal {S}_1 \otimes
\mathbb{I}_b$ has by itself a practical significance, as one can
infer $D_p[\mathcal {S}(\rho)]$ without resorting to the evolution
equation of the state $\rho$ itself, and this simplifies greatly the
assessment of the robustness of the considered correlation measure
against decoherence.

The existence of the relation \eqref{eq6} is also a remarkable
feature of the dynamics of the correlation measures defined in Eq.
\eqref{eq3} which differs from those observed for different
entanglement measures \cite{FL1, FL2,FL3,FLb1,FLb2,FLb3}. There, the
different entanglement measures obey a factorization law for
arbitrary one-sided quantum channel, but $\rho$ is limited to the
pure states. Here, although the depolarizing channel is limited to
be one-sided, the relation \eqref{eq6} holds for arbitrary (pure and
mixed) initial state $\rho$.

If the depolarizing channel acts locally on $b$ or on $ab$ of the
system, with $\rho$ does not satisfy the condition presented in Eq.
\eqref{eq-c1}, a simple relation of the form of Eq. \eqref{eq6}
cannot be derived in general. But for certain specific correlation
measures, one can still obtain some similar relations determining
their dynamics. For instance, the 2-norm measurement-induced
nonlocality for any $(2\times n)$-dimensional state \cite{min} and
the 1-norm measurement-induced nonlocality for any two-qubit state
\cite{min1} had already been derived, from which one can check that
the factorization decay behavior in Eq. \eqref{eq6} always holds.

Moreover, the 2-norm GQD for arbitrary two-qubit state $\rho$ was
given by \cite{gqd}
%%%%%%%%%%%%%%%%%%%%%%%%%%%
\begin{equation}\label{eq-sp25}
 D_2 (\rho) = \frac{1}{4} [||\vec{x}||_2^2 + ||T||_2^2-k_{\max}(K)],
\end{equation}
%%%%%%%%%%%%%%%%%%%%%%%%%%%
where $k_{\max}(K)$ is the largest eigenvalue of $K = \vec{x}
\vec{x}^t + T T^t$. Then, as $\vec{x}'= \vec{x}$ and $T'=q_b T$ when
party $b$ of the system $ab$ traverses the depolarizing channel, we
have
%%%%%%%%%%%%%%%%%%%%%%%%%%%
\begin{equation}\label{eq-sp26}
 D_2[(\mathbb{I}_a\otimes\mathcal {S}_2)\rho]=\frac{1}{4}[||\vec{x}||_2^2 +
                                    |q_b|^2||T||_2^2-k_{\max}(K')],
\end{equation}
%%%%%%%%%%%%%%%%%%%%%%%%%%%
where $K'=\vec{x}\vec{x}^t+|q_b|^2 TT^t$.

By rewriting $K'$ as $K'=|q_b|^2 K+ (1-|q_b|^2)\vec{x}\vec{x}^t$ and then
using the Weyl's theorem, we know that
%%%%%%%%%%%%%%%%%%%%%%%%%%%
\begin{equation}\label{eq-sp27}
  k_{\max}(K') \leq |q_b|^2 k_{\max}(K) + (1-|q_b|^2)k_{\max}(\vec{x}\vec{x}^t),
\end{equation}
%%%%%%%%%%%%%%%%%%%%%%%%%%%
therefore
%%%%%%%%%%%%%%%%%%%%%%%%%%%
\begin{eqnarray}\label{eq-sp28}
 D_2 [(\mathbb{I}_a \otimes\mathcal {S}_2)\rho]
     & \geq & \frac{1}{4} [||\vec{x}||_2^2 + |q_b|^2 ||T||_2^2-|q_b|^2 k_{\max}(K) \nonumber \\
           && -(1-|q_b|^2)k_{\max}(\vec{x}\vec{x}^t)] \nonumber\\
     & = &   \frac{1}{4}|q_b|^2 [||\vec{x}||_2^2+||T||_2^2-k_{\max}(K)] \nonumber \\
           && + \frac{1}{4}(1-|q_b|^2)[||\vec{x}||_2^2-k_{\max}(\vec{x}\vec{x}^t)] \nonumber \\
     & = &  |q_b|^2 D_2 (\rho),
\end{eqnarray}
%%%%%%%%%%%%%%%%%%%%%%%%%%%
where the second equality is due to $||\vec{x}||_2^2 - k_{\max}
(\vec{x} \vec{x}^t) = 0$.

For $\mathcal {S}_1\otimes\mathcal {S}_2$, by replacing $\rho$ in
Eq. \eqref{eq-sp28} with $(\mathcal {S}_1\otimes\mathbb{I}_b)\rho$
and then using Eq. \eqref{eq6}, one can obtain
%%%%%%%%%%%%%%%%%%%%%%%%%%%
\begin{eqnarray}\label{eq17}
 D_2[(\mathcal {S}_1\otimes \mathcal {S}_2)\rho] \geq |q_a q_b|^2 D_2 (\rho).
\end{eqnarray}
%%%%%%%%%%%%%%%%%%%%%%%%%%%

Thus for the 2-norm GQD, a relation similar to that of Eq.
\eqref{eq6} is still satisfied, with only the original equality
being replaced by an inequality.

\subsection{Pauli channels}
The possible actions of the channel on a qubit can be characterized
by at most four independent Kraus operators that are linear
combinations of the identity and the Pauli matrices. We consider
here the Pauli channel $\mathcal {S}_1 $ (while $\mathcal {S}_2$ is
the same as $\mathcal {S}_1$, except that it acts on subsystem $b$)
with the Kraus operators
%%%%%%%%%%%%%%%%%%%%%%%%%%%
\begin{eqnarray}\label{eq-sp32}
 E_0 = \sqrt{\varepsilon_0} \mathbb{I}_2,~
 E_k = \sqrt{\varepsilon_k} \sigma_{k}~ (k = \{1,2,3\}),
\end{eqnarray}
%%%%%%%%%%%%%%%%%%%%%%%%%%%
where $\sum_{\mu=0}^3 E_\mu^\dag E_\mu = \mathbb{I}_2$. This yields
%%%%%%%%%%%%%%%%%%%%%%%%%%%
\begin{eqnarray}\label{eq-sp33}
 \begin{split}
  & \mathcal {S}_1^\dag(\sigma_1) = (\varepsilon_0 + \varepsilon_1 -
                           \varepsilon_2 - \varepsilon_3) \sigma_1,\\
  & \mathcal {S}_1^\dag(\sigma_2) = (\varepsilon_0 - \varepsilon_1 + \varepsilon_2 -
                           \varepsilon_3) \sigma_2,\\
  & \mathcal {S}_1^\dag(\sigma_3) = (\varepsilon_0 - \varepsilon_1 - \varepsilon_2 +
                           \varepsilon_3) \sigma_3.
 \end{split}
\end{eqnarray}
%%%%%%%%%%%%%%%%%%%%%%%%%%%

Then, by supposing $x_k' = q_k x_k$, we obtain
%%%%%%%%%%%%%%%%%%%%%%%%%%%
\begin{equation}\label{eq-sp34}
 \begin{split}
  & \varepsilon_0 = \frac{1}{4}(1 + q_1 + q_2 + q_3), \\
  & \varepsilon_1 = \frac{1}{4}(1 + q_1 - q_2 - q_3), \\
  & \varepsilon_2 = \frac{1}{4}(1 - q_1 + q_2 - q_3), \\
  & \varepsilon_3 = \frac{1}{4}(1 - q_1 - q_2 + q_3).
 \end{split}
\end{equation}
%%%%%%%%%%%%%%%%%%%%%%%%%%%

This result enables us to construct a number of Pauli channels for
which the evolution equations of the geometric correlations are
governed by Eq. \eqref{eq6}. For instance, by choosing $q_i = q_j
\equiv q$ and $q_k = q_0$, with  $i \neq j \neq k$, we obtain the
Pauli channels of the following form
%%%%%%%%%%%%%%%%%%%%%%%%%%%
\begin{eqnarray}\label{eq25}
 \begin{split}
 &E_{0} = \frac{1}{2}\sqrt{1 + q_0 + 2q}\mathbb{I}_2,~~
  E_{i} = \frac{1}{2}\sqrt{1-q_0}\sigma_{i},\\
 &E_{j} = \frac{1}{2}\sqrt{1-q_0}\sigma_{j},~~
  E_{k} = \frac{1}{2}\sqrt{1 + q_0 - 2q}\sigma_{k}.
\end{split}
\end{eqnarray}
%%%%%%%%%%%%%%%%%%%%%%%%%%%

For any chosen values of $i \neq j \neq k$, Eq. \eqref{eq25} gives
the map $\mathcal {S}_1^\dag (\sigma_{i,j}) = q \sigma_{i,j}$, and
$\mathcal {S}_1^\dag (\sigma_{k}) = q_0 \sigma_{k}$. Thus, it
satisfies the conditions listed in Theorems 2 and 4, and the
families of $\rho$ whose correlation dynamics is governed by Eq.
\eqref{eq6} can be found in Eqs. \eqref{eq-c2} and \eqref{eq-c4}.

\begin{figure}
\centering
\resizebox{0.45\textwidth}{!}{%
\includegraphics{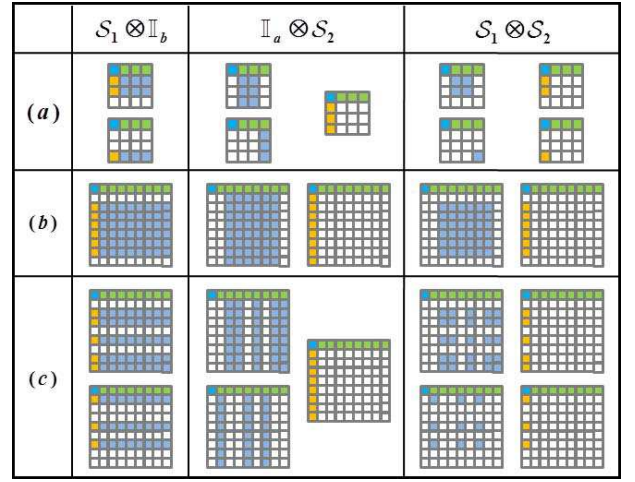}}
% If not, use\vspace{5cm}
% Give the correct figure height in cm
\caption{(Color online) Correlation matrices related to the families
         of $\rho$ for which the evolution equation of $D_p[\mathcal {S}(\rho)]$
         is given by Eq. \eqref{eq6}, with the Kraus operators of the
         channels being given respectively by (a) Eq. \eqref{eq25}
         with $i=1$, $j=2$, and $k=3$, (b) Eq. \eqref{eq32} with
         $k_1=1$, and (c) Eq. \eqref{eq33} with $k_1=1$, $k_2=4$,
         and $k_3=6$. Here, the correlation matrix elements denoted
         by the hollow squares should be zero, while those denoted
         by the color shaded squares can take arbitrary physically
         allowed values.} \label{fig:2}
% Give a unique label
\end{figure}
%%%%%%%%%%%%%%%%%%%%%%%%%%%%%%

For clearness, we summarized in Fig. \ref{fig:2}(a) the above
results, where the correlation matrix is defined as
\begin{eqnarray}\label{eq26}
 R=\left(\begin{array}{cc}
      \frac{1}{\sqrt{d_A d_B}} & y^t \\
                 x             & T   \\
    \end{array}\right),
\end{eqnarray}
%%%%%%%%%%%%%%%%%%%%%%%%%%%
and the square lies in the $i$th row and $j$th column represents the
corresponding correlation matrix elements.

Eq. \eqref{eq25} covers the depolarizing channel ($q_0=q$), and the
bit flip ($i=1$), bit-phase flip ($i=2$), and phase flip ($i=3$)
channels when $q_0=1$. It also enables one to identify those $\rho$
for which the correlations are immune of decay. For example, we have
$\mathcal {S}_1^\dag(\sigma_{1,2}) = q \sigma_{1,2}$ and $\mathcal
{S}_1^\dag(\sigma_3) = \sigma_3$ for the phase flip channel
$\mathcal {S}_1 \otimes \mathbb{I}_2$. Thus, if $x_{1,2}$ and the
first two rows of $T$ in Eq. \eqref{eq2} equal zero, then
$D_p[\mathcal {S}(\rho)]$, $\mathcal {N}_{\rm qt}[\mathcal
{S}(\rho)]$, $\mathcal {F}_{\rm rsp}[\mathcal {S}(\rho)]$, and
$\mathcal {B}_{\rm max} [\mathcal {S}(\rho)]$ will do not decay with
time. Similar results can be obtained for the bit flip and bit-phase
flip channels.

The channels of Eq. \eqref{eq25} are unital, i.e., $\mathcal {S}_1
(\mathbb{I}_2 /2) = \mathbb{I}_2/2$. We now further consider an
nonunital channel, i.e., the generalized amplitude damping channel
with
%%%%%%%%%%%%%%%%%%%%%%%%%%%
\begin{equation}\label{eq27}
 \begin{split}
  & E_{0} = \frac{1}{2}\sqrt{\eta_{\bar{n}}} [(1+q)\mathbb{I}_2 - (1-q)\sigma_3],\\
  & E_{1} = \frac{1}{2}\sqrt{\eta_{\bar{n}'}}[(1+q)\mathbb{I}_2 + (1-q)\sigma_3],\\
  & E_{2} = \sqrt{\eta_{\bar{n}}} \sqrt{1-q^2}\sigma_{-},\\
  & E_{3} = \sqrt{\eta_{\bar{n}'}}\sqrt{1-q^2}\sigma_{+},
 \end{split}
\end{equation}
%%%%%%%%%%%%%%%%%%%%%%%%%%%
where $\sigma_{\pm}= (\sigma_1 \pm i\sigma_2)/2$,
$\eta_{\bar{n}'}=1-\eta_{\bar{n}}$, and $\eta_{\bar{n}}$ is a
parameter determined by the average thermal photons $\bar{n}$ in the
reservoir \cite{thermal}. When $\eta_{\bar{n}}=1$, one recovers the
zero temperature reservoir, which has been discussed extensively
with the Lorentzian, or the sub-Ohmic, Ohmic, and super-Ohmic type
spectral densities \cite{dyn4}.

This generalized amplitude damping channel gives the map $\mathcal
{S}_1^\dag(\sigma_{1,2}) = q \sigma_{1,2}$, and $\mathcal
{S}_1^\dag(\sigma_3) = [1+2\eta_{\bar{n}}(q^2-1)] \mathbb{I}_2 -
2q^2 |1\rangle\langle 1|$, with $|1\rangle$ being the lower energy
state of $\sigma_3$. As a result, it also satisfies the conditions
listed in Theorems 2 and 4, and thus the families of $\rho$ for
which the evolution equations of $D_p[\mathcal {S}(\rho)]$ and
$\tilde{D}_p[\mathcal {S}(\rho)]$ obey a factorization decay
behavior can be obtained directly by using Eqs. \eqref{eq-c2} and
\eqref{eq-c4}.

\subsection{Gell-Mann channels}
The Gell-Mann matrices are of the following form
%%%%%%%%%%%%%%%%%%%%%%%%%%%
\begin{equation}\label{eq-sp35}
 \begin{split}
  &\lambda_1 = \left(\begin{array}{ccc}
               0  & 1 & 0 \\
               1  & 0 & 0 \\
               0  & 0 & 0
               \end{array}\right),~
  \lambda_2 = \left(\begin{array}{ccc}
               0  & -i & 0 \\
               i  & 0  & 0 \\
               0  & 0  & 0
              \end{array}\right),\\
  &\lambda_3 = \left(\begin{array}{ccc}
               1  & 0  & 0 \\
               0  & -1 & 0 \\
               0  & 0  & 0
               \end{array}\right),~
  \lambda_4 = \left(\begin{array}{ccc}
               0  & 0 & 1 \\
               0  & 0 & 0 \\
               1  & 0 & 0
              \end{array}\right),\\
  &\lambda_5 = \left(\begin{array}{ccc}
               0  & 0 & -i \\
               0  & 0 & 0 \\
               i  & 0 & 0
               \end{array}\right),~
  \lambda_6 = \left(\begin{array}{ccc}
               0  & 0 & 0 \\
               0  & 0 & 1 \\
               0  & 1 & 0
              \end{array}\right), \\
  &\lambda_7 = \left(\begin{array}{ccc}
               0  & 0 & 0 \\
               0  & 0 & -i \\
               0  & i & 0
               \end{array}\right),~
  \lambda_8 = \frac{1}{\sqrt{3}}\left(\begin{array}{ccc}
               1  & 0 & 0 \\
               0  & 1 & 0 \\
               0  & 0 & -2
              \end{array}\right).
 \end{split}
\end{equation}
%%%%%%%%%%%%%%%%%%%%%%%%%%%

The possible actions of the channel on a qutrit can be described by
the Kraus operators that are linear combinations of the identity and
the Gell-Mann matrices, and for convenience of presentation, we call
them Gell-Mann channels. Similar as the qubit system, we consider
here the representative class of $\mathcal {S}_1$ (with $\mathcal
{S}_2$ being the same as $\mathcal {S}_1$, except that it acts on
subsystem $b$) with the Kraus operators being given by
%%%%%%%%%%%%%%%%%%%%%%%%%%%
\begin{equation}\label{eq-sp36}
 E_0=\sqrt{\varepsilon_0}\mathbb{I}_3,~
 E_k=\sqrt{\varepsilon_k}\lambda_k~(k=\{1,2,\ldots,8\}),
\end{equation}
%%%%%%%%%%%%%%%%%%%%%%%%%%%
where $\sum_{\mu=0}^8 E_\mu^\dag E_\mu = \mathbb{I}_3$.

Then, we suppose $\mathcal {S}_1^\dag(\lambda_k)=q_k \lambda_k$ for
the purpose of finding the channel under the action of which the
evolution equations of the correlation measures obey a factorization
decay behavior. After a straightforward calculation, we obtain that
the parameters $q_k$ must satisfy the following relations
%%%%%%%%%%%%%%%%%%%%%%%%%%%
\begin{equation}\label{eq-sp37}
 \begin{split}
  & q_1 + q_2 + q_3 = q_6 + q_7 + q_8, \\
  & q_4 + q_5 = q_6 + q_7,
 \end{split}
\end{equation}
%%%%%%%%%%%%%%%%%%%%%%%%%%%
under which we have
%%%%%%%%%%%%%%%%%%%%%%%%%%%
\begin{eqnarray}\label{eq-sp38}
 \begin{split}
  & \varepsilon_0 = \frac{1}{9}(1+3q_6+3q_7+2q_8), \\
  & \varepsilon_1 = \frac{1}{12}(2+6q_1-3q_6-3q_7-2q_8), \\
  & \varepsilon_2 = \frac{1}{12}(2+6q_2-3q_6-3q_7-2q_8), \\
  & \varepsilon_3 = \frac{1}{12}(2+6q_3-3q_6-3q_7-2q_8), \\
  & \varepsilon_4 = \frac{1}{12}(2 + 3q_4 - 3q_5-2q_8), \\
  & \varepsilon_5 = \frac{1}{12}(2 - 3q_4 + 3q_5-2q_8), \\
  & \varepsilon_6 = \frac{1}{12}(2 + 3q_6 - 3q_7-2q_8), \\
  & \varepsilon_7 = \frac{1}{12}(2 - 3q_6 + 3q_7-2q_8), \\
  & \varepsilon_8 = \frac{1}{12}(2-3q_6-3q_7+4q_8).
 \end{split}
\end{eqnarray}
%%%%%%%%%%%%%%%%%%%%%%%%%%%

The depolarizing channel for a qutrit can be considered as a special
case of Eq. \eqref{eq-sp36}, which corresponds to $q_k = q$ for all
$k=\{1,2,\ldots,8\}$. Moreover, by choosing $q_{k_1,8} = 1$ ($k_1 =
1$, $2$, or $3$), and $q_k = q$ for all $k \notin \{k_1,8\}$, we
present the class of the Gell-Mann channels with
%%%%%%%%%%%%%%%%%%%%%%%%%%%
\begin{eqnarray}\label{eq32}
 \begin{split}
  &  E_0 = \sqrt{\frac{1+2q}{3}} \mathbb{I}_3,~
     E_1 = \sqrt{\frac{1-q}{2}} \lambda_{k_1},\\
  &  E_2 = \sqrt{\frac{1-q}{2}} \lambda_{8},
 \end{split}
\end{eqnarray}
%%%%%%%%%%%%%%%%%%%%%%%%%%%
which gives $\mathcal {S}_1^\dag(\lambda_k) = \lambda_k$ for $k \in
\{k_1,8\}$, and $\mathcal {S}_1^\dag(\lambda_k) = q \lambda_k$
otherwise. Clearly, it satisfies the conditions listed in Theorems 2
and 4, and thus the families of $\rho$ for which the evolution
equations of the considered correlations are governed by Eq.
\eqref{eq6} can be found in Eqs. \eqref{eq-c2} and \eqref{eq-c4}.
See Fig. \ref{fig:2}(b) for the related correlation matrices.

By choosing $q_{k_1,k_2,k_3}=q$ (here $k_1=1$, $2$, or $3$, $k_2=4$
or $5$, and $k_3=6$ or $7$), and $q_{k}=3q-2$ for all $k\notin
\{k_1,k_2,k_3\}$, we present here another class of the Gell-Mann
channels described by the Kraus operators
%%%%%%%%%%%%%%%%%%%%%%%%%%%
\begin{eqnarray}\label{eq33}
 \begin{split}
  & E_0 = (2q-1) \mathbb{I}_3,~ E_{1}=(1-q)\lambda_{k_1},\\
  & E_{3}=(1-q)\lambda_{k_2},~  E_{3}=(1-q)\lambda_{k_3},
 \end{split}
\end{eqnarray}
%%%%%%%%%%%%%%%%%%%%%%%%%%%
which gives the map $\mathcal {S}_1^\dag(\lambda_k)= q \lambda_k$
for $k\in \{k_1,k_2, k_3\}$, and $\mathcal {S}_1^\dag(\lambda_k)
=(3q-2)\lambda_k$ otherwise. Thus, the families of $\rho$ for which
$D_p[\mathcal {S}(\rho)]$ and $\tilde{D}_p[\mathcal {S}(\rho)]$ obey
the factorization decay behavior can be obtained from Eqs.
\eqref{eq-c2} and \eqref{eq-c4}, respectively. See Fig.
\ref{fig:2}(c) for the related correlation matrices.

\section{Summary and discussion}\label{sec:5}
In summary, we have investigated the evolution equations for a
series of geometric correlation measures. These measures are all
related with certain forms of the Schatten $p$-norm. We have
established a simple relation that determines the evolution of these
correlation measures. This relation is of central relevance for
assessing the robustness of the related correlations which might be
the resource of QIP tasks such as the input-output gate operation in
sequential quantum computing and the experimental generation of
quantum correlated resources in noisy environments. Moreover, as
$\mathcal {S}(\rho)$ may represents the action of environment, of
measurements, or of both on $\rho$, and the factorization decay
behavior applies to various systems, various discord-like
correlation measures, and even further various figures of merit
associated with quantum teleportation, remote state preparation, and
Bell-inequality violation, the criteria provided in this work are
also useful for easing the evaluation of correlations in these
tasks. A deep exploration of these relations might also be related
with the structure of entanglement spectrum in describing topology
of band structures of many-body systems \cite{haldane,fan1,fan2}. We
hope these results may shed some lights on understanding the essence
of quantum correlations and their applications in QIP and condensed
matter physics.

\section*{ACKNOWLEDGMENTS}
The authors thank useful discussions with Florian Mintert. This work
was supported by NSFC (11205121, 11175248), the ``973'' program
(2010CB922904), and NSF of Shaanxi Province (2014JM1008).

\newcommand{\PRL}{Phys. Rev. Lett. }
\newcommand{\RMP}{Rev. Mod. Phys. }
\newcommand{\PRA}{Phys. Rev. A }
\newcommand{\PRB}{Phys. Rev. B }
\newcommand{\NJP}{New J. Phys. }
\newcommand{\JPA}{J. Phys. A }
\newcommand{\JPB}{J. Phys. B }
\newcommand{\PLA}{Phys. Lett. A }
\newcommand{\NP}{Nat. Phys. }
\newcommand{\NC}{Nat. Commun. }
\newcommand{\QIC}{Quantum Inf. Comput. }
\newcommand{\QIP}{Quantum Inf. Process. }
\newcommand{\EPJD}{Eur. Phys. J. D }

%

% BibTeX users please use
% \bibliographystyle{}
% \bibliography{}

\end{document}